\documentstyle[12pt]{article}
\jot = 1.5ex\catcode`\@=11\renewcommand{\thefootnote}{\fnsymbol{footnote}}

\def\be{\begin{equation}}
\def\ee{\end{equation}}
\def\l{\label}
\def\F{{\cal F}}

\def\ldl{\Lambda\partial_\Lambda}

\begin{document}\begin{titlepage}

\hfill{DFPD/96/TH/25}

\hfill{hep-th/9605090}

\vspace{1cm}

\centerline{\large{\bf NONPERTURBATIVE RELATIONS IN}}

\vspace{0.5cm}

{\centerline{\large{\bf N=2 SUSY YANG-MILLS AND WDVV EQUATION}}}

\vspace{1.5cm}

{\centerline{\sc Giulio BONELLI and Marco 
MATONE\footnote[3]{Partly
supported by the European Community Research
Programme {\it Gauge Theories, applied supersymmetry and quantum 
gravity}, contract SC1-CT92-0789}}}

\vspace{0.8cm}

\centerline{\it Department of Physics ``G. Galilei'' - Istituto Nazionale di 
Fisica Nucleare}
\centerline{\it University of Padova}
\centerline{\it Via Marzolo, 8 - 35131 Padova, Italy}

\centerline{bonelli@padova.infn.it   matone@padova.infn.it}

\vspace{2cm}

\centerline{\bf ABSTRACT}

\vspace{0.6cm}

\noindent
We find the nonperturbative relation between $\langle {\rm tr}\, 
\phi^2\rangle$, $\langle {\rm tr}\, \phi^3\rangle$
the prepotential ${\cal F}$ and the vevs $\langle \phi_i\rangle$
 in $N=2$ supersymmetric Yang-Mills theories with gauge group $SU(3)$.
Nonlinear differential equations 
for ${\cal F}$ including the 
Witten -- Dijkgraaf -- Verlinde -- Verlinde equation are obtained.
This indicates that $N=2$ SYM theories are essentially topological field 
theories and that should be seen as low-energy limit of some topological 
string theory.
Furthermore, we construct relevant modular invariant quantities, 
derive canonical relations between the periods
and investigate the structure of the beta function by giving
its explicit form in the moduli coordinates. 
In doing this we discuss
the uniformization problem for the quantum moduli space.
 The method we propose can be generalized to $N=2$ supersymmetric 
Yang-Mills theories with higher rank gauge groups.

\end{titlepage}

\newpage
\setcounter{footnote}{0}
\renewcommand{\thefootnote}{\arabic{footnote}}

\noindent
{\bf 1.} Seiberg-Witten exact results about $N=2$ SUSY Yang-Mills \cite{SW1}
concern the low-energy Wilsonian effective action with at most two derivatives 
and four fermions. These terms are completely described by the so-called 
prepotential ${\cal F}$ whose most important property is holomorphicity 
\cite{Gates}. Furthermore, it has been shown in \cite{S} that ${\cal F}$ gets 
perturbative contributions only up to one-loop. Higher-order terms in the 
asymptotic expansion come as instanton contribution implicitly determined in 
\cite{SW1}.

In \cite{m}, where a method to invert functions was proposed, it has been 
derived a nonperturbative equation which relates in a simple way the 
prepotential and the vevs of the scalar fields. In \cite{BM}, proving a 
conjecture in \cite{STY}, it has been shown that the above relation underlies 
the nonperturbative Renormalization Group Equation and the exact expression 
for the beta function in the $SU(2)$ case has been obtained.
The problem of extending these results to the case of higher rank groups is a 
nontrivial task. An important step in this direction is the result in 
\cite{STY}\cite{EY} where the nonperturbative relation in \cite{m} has been 
generalized. However, it remains the problem of finding the nonperturbative 
relations between $\langle{\rm tr}\, \phi^k\rangle$ for $k>2$ and the 
prepotential.
Also, one should find a set of equations for ${\cal F}$
in a similar way to the $SU(2)$ case \cite{m}. 

In this paper we will solve these problems for the $SU(3)$ case.
In particular, we will find a complete set of non-linear 
differential equations completely characterizing the prepotential including the
Witten --  Dijkgraaf -- Verlinde  -- Verlinde (WDVV) 
equation \cite{WDVV}.
This indicates that $N=2$ SYM theories are essentially topological field 
theories and that should be seen as low-energy limit of some topological 
string theory.

Furthermore, we introduce a set of modular invariant quantities
 which will be useful to find the relation between
$\langle{\rm tr}\, \phi^k\rangle$ and ${\cal F}$
and to formulate canonical relations between the periods.
We also investigate the structure of the beta function and give its 
explicit form in the moduli coordinates.

\vspace{1cm}

\noindent
{\bf 2.} 
The Seiberg-Witten curve for $SU(n)$, 
has been found in \cite{KLYT} for $n=3$ and generalized to arbitrary
$n$ in \cite{AF}.
Let us denote by $a^i= \langle \phi^i\rangle$ and $a_i^D=\langle 
\phi_i^D\rangle=\partial{\cal F}/\partial a^i$ the vevs of the scalar component
of the chiral superfield and its dual. The effective couplings are given by
$\tau_{ij}=\partial^2 {\cal F}/\partial a^i\partial a^j$. We also
set $u^2\equiv u=\langle {\rm tr\, \phi^2}\rangle$, $u^3\equiv v=\langle{\rm tr
\, \phi^3}\rangle$ and $\partial_k\equiv \partial/\partial a^k$, 
$\partial_\alpha\equiv \partial/\partial u^\alpha$. 
Our starting point are the 
reduced Picard -- Fuchs  equations (RPFE's) for $SU(3)$ introduced in \cite{KLT}
\begin{equation}
{\cal L}_\beta \left(\begin{array}{c} a_i^D \\ a^i \end{array}\right)=0,
\qquad \beta=2,3,
\label{x1}\end{equation}
where
$$
{\cal L}_2={1\over u} P(u,v,\Lambda)\partial_u^2+{\cal L},\qquad \qquad 
{\cal L}_3={1\over 3} P(u,v,\Lambda)\partial_v^2+{\cal L},
$$
\begin{equation}
P(u,v,\Lambda)=27(v^2-\Lambda^6)+4u^3,\qquad 
{\cal L}=12uv\partial_u\partial_v+3v\partial_v+1.
\label{x2}\end{equation}

Let us recall some transformation properties under 
the action of $Sp(2n-2,{\bf Z})$ which hold for the $SU(n)$ case.
We have
$$
\left(\begin{array}{c} a^D\\ a 
\end{array}\right)\longrightarrow
\left(\begin{array}{c} \tilde 
a^D\\ \tilde a \end{array}\right)=
\left(\begin{array}{c}A\\C
\end{array}\begin{array}{cc}B\\D\end{array}\right)
\left(\begin{array}{c} a^D\\ a \end{array}\right).
$$
Integrating $\tilde a^D=A a^D+Ba=\partial \widetilde{\cal F}(\tilde a)/
\partial \tilde a$ yields \cite{dWKLL,m,STY}
$$
\widetilde {\cal F}(\tilde a)=
{\cal F}(a)+{1\over 2} a^D C^tA a^D+{1\over 2} a B^t D a+
aB^tC a^D.
$$

We now re-write the equations (\ref{x1})
as non-linear differential equations with respect 
to the $a^i$-coordinates. 
To this end it is convenient to introduce the following notation
$$
{\cal U}=u_2^2\partial_{11}-2u_1u_2\partial_{12}+u_1^2\partial_{22},\qquad
{\cal V}=v_2^2\partial_{11}-2v_1v_2\partial_{12}+v_1^2\partial_{22},
$$
$$
{\cal C}=(u_1v_2+v_1u_2)\partial_{12}-u_2v_2\partial_{11}-u_1v_1\partial_{22},
\qquad D=u_1v_2-u_2v_1,
$$
where $\partial_{i_1...i_n}
\equiv {\partial^n/ \partial a^{i_1}...\partial a^{i_n}}$,
$u_i\equiv \partial_{i} u$ and
$v_i\equiv \partial_{i} v$.
In computing the inversion of Eqs.(\ref{x1})
there are terms which simplify because
the $a^i$ are solutions of the
RPFE's themselves. We have 
\begin{equation}
\left[12uv {\cal C}
+{1\over 3}P(u,v,\Lambda){\cal U}+D^2(1-a^i\partial_i)\right]{\cal F}_{l}=0
=\left[12uv {\cal C}+{1\over u} P(u,v,\Lambda)
{\cal V}+D^2(1-a^i\partial_i)\right]{\cal F}_{l},
\label{x3}\end{equation}
where $l=1,2$ and ${\cal F}_{i_1...i_n}\equiv 
\partial_{i_1...i_n}{\cal F}$.
Note that $D$ is the Jacobian of $(u,v)\to (a^1,a^2)$
and therefore generally non-vanishing.
Subtracting the LHS from the RHS of Eqs.(\ref{x3}), we obtain
\begin{equation}
A_l\equiv x_{11}{\cal F}_{22l}+x_{22}{\cal F}_{11l}-2x_{12}
{\cal F}_{12l}=0, 
\label{x5}\end{equation}
where $l=1,2$ and
\be
x_{ij}=3v_iv_j-uu_iu_j.
\l{nnnnn}\ee
We stress that Eqs.(\ref{x5}) have been obtained from the Eqs.(\ref{x1}).
Therefore, since $(\tilde a^D_i,\tilde a^i)$ are still solutions of 
Eqs.(\ref{x1}), it follows that Eqs.(\ref{x5})
are modular invariant by construction. 

\vspace{1cm}

\noindent
{\bf 3.} 
On general grounds it seems that the Picard-Fuchs equations are
related to the WDVV equation. 
The above construction allows us to show that
this is actually true for the Picard-Fuchs equations
arising in $N=2$ supersymmetric Yang-Mills theory with 
gauge group $SU(3)$. 
Actually, a suitable linear combination
of the equations $A_l=0$, namely
$$
A_1\left(y_{22} {\cal F}_{112} -2y_{12} {\cal F}_{122}
+y_{11}{\cal F}_{222}\right)-A_2
\left(-2y_{12} {\cal F}_{112} +y_{11} {\cal F}_{122}
+y_{22}{\cal F}_{111}\right)=0,
$$
where $y_{jk}$ are arbitrary parameters, can be written
in the WDVV form
\be
{\cal F}_{ikl}\eta^{lm}{\cal F}_{mnj}=
{\cal F}_{jkl}\eta^{lm}{\cal F}_{mni},
\label{wdvvv}\ee
for $i,j,k,n=1,2$, where
\be
\eta^{lm}=
\left(\begin{array}{c}2x_{22}y_{12}-2x_{12}y_{22}\\x_{11}y_{22}-
x_{22}y_{11}
\end{array}\begin{array}{cc}x_{11}y_{22}-x_{22}y_{11}
\\2x_{12}y_{11}-2x_{11}y_{12}\end{array}\right).
\label{hfgty}\ee
We observe that for each choice of the metric, that is of the parameters 
$y_{jk}$, there is only one nontrivial equation in 
(\ref{wdvvv}) which can be re-written as
\be
\eta^{11}\Theta_{11}+2\eta^{12}\Theta_{12}+\eta^{22}\Theta_{22}=0,
\l{oiuhd}\ee
where 
$$
\Theta_{ij}={1\over 2}\left(
{\cal F}_{11i}{\cal F}_{22j}+{\cal F}_{11j}{\cal F}_{22i}\right)-
{\cal F}_{12i}{\cal F}_{12j},
$$
which satisfy the identity
\be
2{\cal F}_{12l}\Theta_{12}={\cal F}_{22l}\Theta_{11}
+{\cal F}_{11l}\Theta_{22},\qquad l=1,2.
\l{osita}\ee

\vspace{1cm}

\noindent
{\bf 4.} Let us introduce some modular invariant quantities which will be used 
later on. We set
\begin{equation}
I^{\;\;  \gamma}_\beta=(\partial_k z)
{\left(\partial_\beta\tau\right)^{-1}}^{kl}
\partial_l u^\gamma, 
\label{hfgy}\end{equation}
where $\beta,\gamma=2,3$ and $z$ is the modular invariant
$$
z=a^i\partial_i\F-2\F.
$$
Another set of modular invariants which will be useful are 
\begin{equation}
v_{(\beta)}^{\;\; \alpha}= 
I^{\;\; \gamma}_\beta(\partial_\gamma\partial_k u^\alpha)
 \partial_\beta a^k 
+ a^k\partial_k u^\alpha, \qquad \alpha,\beta=2,3.
\label{oifjp}\end{equation}

There is an interesting structure underlying
these invariants. Namely, introducing the brackets 
\be
\left\{X,Y\right\}_{(\beta)}\equiv
\partial_iX{\left(\partial_\beta\tau\right)^{-1}}^{ij}\partial_\beta
\partial_jY-
\partial_iY{\left(\partial_\beta\tau\right)^{-1}}^{ij}\partial_\beta
\partial_jX,
\l{brackets}\ee
the vector field components $v_{(\beta)}^\alpha$ can be expressed
in the form
\be
v_{(\beta)}^\alpha=\{u^\alpha,z\}_{(\beta)}.
\l{annnnn}\ee
Furthermore, the periods satisfy the following canonical relations
\be
\{a^i,a^j\}_{(\beta)}=0,\qquad
\{a^D_i,a^D_j\}_{(\beta)}=0,\quad
\{a^i,a^D_j\}_{(\beta)}=\delta^i_j.
\l{ttttt}\ee

In order to extract the differential equations for ${\cal F}$,
we re-write the operators in (\ref{x2}) in the following general form 
\begin{equation}
{\cal L}_{\beta}=\xi_{(\beta)}\partial_{\beta}
+\eta_{(\beta)}+1,
\label{genoveffa11}\end{equation}
where $\xi_{(\beta)}=\xi_{(\beta)}^{\;\; \alpha}\partial_\alpha=
\xi_{(\beta)}^{\;\; i}\partial_i$ and
$\eta_{(\beta)}=\eta_{(\beta)}^{\;\; \alpha}\partial_\alpha=
\eta_{(\beta)}^{\;\; i}\partial_i$
are vector fields. Considering the action of ${\cal L}_\beta$ on $fg$ with $f$ 
and $g$ arbitrary functions, we have 
$$
{\cal L}_\beta fg= g {\cal L}_\beta f +
f {\cal L}_\beta g -fg + \partial_\beta f \xi_{(\beta)} g +
\xi_{(\beta)}f\partial_\beta g,
$$
and by Eqs.(\ref{x1}) 
\be
{\cal L}_\beta (a^ia_i^D-2{\cal F})=
a^ia_i^D-2{\cal F},
\label{ookiwlpk}\ee
that is ${\cal L}_\beta z=z$.
Note that in (\ref{genoveffa11}), as in (\ref{x2}), for each value of $\beta$ 
the second-order derivative terms contain always at least 
one $\partial_\beta$ (note that 
Eq.(\ref{ookiwlpk}) is independent from this peculiarity).

In order to find $\xi_{(\beta)}$ and $\eta_{(\beta)}$,
we impose that the operators defined in (\ref{genoveffa11}) satisfy
(\ref{x1}). From the lower components of (\ref{x1}) we obtain
$\eta_{(\beta)}^{\;\; i}=-a^i-\xi_{(\beta)} \partial_\beta a^i$,
which substituted in the upper components of (\ref{x1}), yields
$\xi_{(\beta)}^{\;\; i}
=(\partial_k z){\left(\partial_\beta\tau\right)^{-1}}^{ki}$.
Therefore 
\be
{\cal L}_{\beta}=I^{\;\; \gamma}_\beta
\partial_{\gamma}\partial_\beta-
v_{(\beta)}^{\;\; \gamma}\partial_{\gamma}+1.
\label{genoveffa}\ee

Comparing (\ref{genoveffa}) with (\ref{x2}) we obtain a complete set of 
non-linear differential equations for the prepotential and 
its exact relation with the moduli coordinates, namely
\be
v_{(2)}^{\;\; 2}=0=v_{(3)}^{\;\; 2}, 
\l{cccc1}\ee
\be
v_{(2)}^{\;\; 3}=-3v=v_{(3)}^{\;\; 3},
\l{cccc2}\ee
\be
I^{\;\; 3}_2=12uv=I^{\;\; 2}_3,
\l{cccc3}\ee
\be
uI^{\;\; 2}_2=P(u,v,\Lambda)=3I^{\;\; 3}_3.
\l{cccc4}\ee

The same procedure introduced above when
applied to the $SU(2)$ case gives $u=Az+B$ (where the constant $A$, 
which turns out to be ${\pi/2i}$, is fixed by asymptotic analysis whereas the 
constant $B$ turns out to be zero using the recursion relations which 
follow from the inversion of the reduced uniformizing equation) and
\begin{equation}
I^{\;\; 2}_2=4(u^2-\Lambda^4),
\label{oifchw}\end{equation}
which is the equation for the prepotential obtained in \cite{m,mm}.

\vspace{1cm}

\noindent
{\bf 5.} Let us define the modular invariant 1-form
\be
W=\left(a^i\partial_\beta a^D_i-a_i^D
\partial_\beta a^i\right)du^\beta=dz,
\label{ooijfdo}\ee
which, due to the existence of the prepotential, is closed, i.e. $dW=0$.
Substituting in (\ref{ooijfdo}) the expression of the periods in terms of 
Appell's $F_4$ functions obtained in \cite{KLT}, we obtain $W={2i\over\pi}du$.
On the other hand, by (\ref{ookiwlpk}) it follows that
 the components of $W=W_\beta du^\beta$ 
satisfy the linear differential equations
\be
\xi_{(\beta)}W_\beta=v_{(\beta)}^{\;\; \alpha} W_\alpha,\qquad 
\beta=2,3,
\label{hfggyt}\ee
which are satisfied by 
$W_2={2i\over \pi}$, $W_3=0$.
Therefore, we have $z={2i\over \pi} u$, that is
\be
u= {\pi i}\left({\cal F}-{a^i\over 2}\partial_i\F\right),
\l{idulk}\ee
in agreement with \cite{STY,EY}. We stress that a possible 
non-vanishing additive constant in (\ref{idulk}) cannot 
be excluded at this stage. Note that by (\ref{idulk}), thanks to 
(\ref{annnnn}), Eq.(\ref{cccc1}) is identically satisfied.

We now use Eq.(\ref{x5}) to face the problem of finding the explicit relation 
between $v$ and the prepotential. 
On general grounds it can be shown that the properties of special 
geometry imply that $\Theta_{ii}\ne 0$ \cite{CerDaFe}.
By (\ref{osita}) the general solution of Eq.(\ref{x5}) is given by
\be
x_{ij}={\rho}\Theta_{ij},
\l{oihdpl}\ee
where $\rho$ is determined by the compatibility condition
$(3v_1v_2)^2=(3v_1^2)(3v_2^2)$ applied
to (\ref{nnnnn}) and (\ref{oihdpl}),  that is
\be
\rho^2\Delta=u\rho(\Theta_{11}u_2^2+\Theta_{22}u_1^2-2u_1u_2\Theta_{12}),
\l{1w}\ee
where $\Delta=\Theta_{12}^2-\Theta_{11}\Theta_{22}$.
Notice that
 $\rho\ne 0$ otherwise $3v_iv_j= uu_iu_j$ which implies $D=0$.
Since 
$\rho^2\Delta=x_{12}^2-x_{11}x_{22}=3uD^2$, we have
$\Delta\ne 0$, so that (\ref{1w}) can be solved as 
$$
\rho=\Delta^{-1}u(\Theta_{11}u_2^2+\Theta_{22}u_1^2-2u_1u_2\Theta_{12}),
$$
which implies 
\be
\left(\begin{array}{c} v_1\\ v_2 
\end{array}\right)=
\epsilon
\sqrt{{u\over3\Delta}}
\left(\begin{array}{c}\Theta_{12}\\ \Theta_{22}
\end{array}\begin{array}{cc}-\Theta_{11}\\-\Theta_{12}\end{array}\right)
\left(\begin{array}{c} u_1\\ u_2 \end{array}\right),
\l{hgjhhg}\ee
where $\epsilon=\pm 1$ and the relative sign between $v_1$ and $v_2$
has been fixed by $x_{12}=\rho\Theta_{12}$.
Observe that we can set $\epsilon=1$ by a
suitable transformation on the moduli variables (we use the 
fact that the RPFE's are invariant under the transformations
$u\to e^{2i\pi k/3}u$ and $v\to -v$). 

In order to find $v$ we first explicitly evaluate the $I^{\;\; 
\gamma}_\beta$ invariants
in terms of $u_i$, $v_i$ and $\F_{ijk}$. Substituting 
(\ref{hgjhhg}) we obtain the relation between  $v$ and ${\cal F}$
and non-linear differential equations for the prepotential as well.

The essential point is that by (\ref{idulk}) and (\ref{cccc3})
we have
\be
v={I^{\;\;  2}_3 I^{\;\;  2}_2\over 
36 I^{\;\;  3}_3}.
\label{vesplicita}\ee
On the other hand
(\ref{hfgy}) can be written as
$$
I^{\;\;  \alpha}_2={2i\over \pi}D\left(v_2^2\Theta_{11}+
v_1^2\Theta_{22}-2v_1v_2\Theta_{12}\right)^{-1}(v_2g_1^\alpha 
-v_1g_2^\alpha),
$$
and
$$
I^{\;\;  \alpha}_3={2i\over \pi}D\left(u_2^2\Theta_{11}+
u_1^2\Theta_{22}-2u_1u_2\Theta_{12}\right)^{-1}
(u_1g_2^\alpha 
-u_2g_1^\alpha),
$$
where
$$
g_i^\alpha=
u_2u_2^\alpha{\cal F}_{11i}+
u_1u_1^\alpha{\cal F}_{22i}-(u_1u_2^\alpha+u_2u_1^\alpha){\cal 
F}_{12i}.
$$

By (\ref{idulk}) and (\ref{hgjhhg}) the $I^{\;\;  \alpha}_\beta$'s
are explicitly known in terms of $a^i$ and the prepotential. It follows
that (\ref{vesplicita}) solves the problem of finding the relation 
between $v$ and ${\cal F}$.

By (\ref{annnnn}), we can re-write Eqs.(\ref{cccc2})
in the form
\be
\{u,v\}_{(\beta)}={6i\over \pi}v,\qquad \beta=2,3,
\l{1q}\ee
which by (\ref{vesplicita}) are two non-linear differential equations for 
${\cal F}$, that together with the two-parameters WDVV equations 
correspond to the four non-linear differential equations (\ref{x3}).

\vspace{1cm}

\noindent
{\bf 6.} Let us now consider the modular properties of the prepotential 
and its homogeneity.
The fact that $\tau_{ij}$ is dimensionless implies that
\be
(\ldl + \Delta_{u,v})\tau_{ij}=0,
\l{oidx1}\ee
where $\Delta_{u,v}= 
2u\partial_u+3v\partial_v$ is the scaling invariant vector field. 
Let $\xi$ be an arbitrary modular 
invariant vector field. We have
$$
\xi\tau\to {\xi\tilde\tau}=(\tau C^t+D^t)^{-1}\xi\tau
(C\tau +D)^{-1},
$$
which implies that (\ref{oidx1}) is a modular invariant equation.
We also have $$
(\ldl + \Delta_{u,v})a^i=a^i,
\qquad
(\ldl + \Delta_{u,v})a^D_i=a^D_i,
$$
which are are compatible with a pseudo-homogeneity
of degree 2 for the prepotential
$$
(\ldl + \Delta_{u,v})\F=2\F+\Lambda^2\cdot{\rm const},
$$
In our case the semiclassical analysis gives const=0.

Let us now discuss in the $SU(3)$ case 
the uniformization mechanism which generalizes the
structure underlying the $SU(2)$ case.
The structure of the covering of the quantum moduli space 
${\cal M}_{SU(3)}$ is encoded in the properties of the Appell's functions. 
The Appell system $F_4$ is a two-dimensional generalization of the 
hypergeometric system also endowed with algebraic relations involving 
the functions and their derivatives.

It is known \cite{KLT} that the period matrix $\tau_{ij}$ is a rational 
combination of Appell's functions. By (\ref{oidx1}), the 
dependence on $u,v$ and $\Lambda$ is of the form
\be
\tau=\tau(u/\Lambda^2,v/\Lambda^3).
\l{poipoi}\ee
Therefore the $\tau$-space is a subvariety ${\cal S}$ of the genus 2 Siegel
upper-half space of complex codimension 1 which covers the 
quantum moduli space.
${\cal S}$ can be characterized as 
${\tt s}(\tau)=0$, where the structure of ${\tt s}$ 
is related to the equations satisfied by the prepotential.

Let $M_{SU(3)}\subset Sp(4,{\bf Z})$ be the monodromy group of $N=2$
SYM with gauge group $SU(3)$ \cite{KLT}. The above remarks imply that 
the Picard-Fuchs equations, from which Eqs.(\ref{x1}) are derived, are the 
uniformization equations for the quantum moduli space. Therefore
\be
{\cal M}_{SU(3)}\cong {{\cal S}/M_{SU(3)}}.
\l{lampesco}\ee
The polymorphic matrix function $\tau$ is the inverse covering with 
$M_{SU(3)}$-monodromy. 
Let
$$
u/\Lambda^2={\tt u}(\tau),\quad
v/\Lambda^3={\tt v}(\tau),\quad
\tau\in {\cal S},
$$
be the covering map. From the above data we now derive the beta function of 
the theory.

Let us consider the following equations
$$
0=\ldl {\tt s}(\tau)=\Sigma(\beta){\tt s}(\tau),
$$
$$
0=\ldl u=\Lambda^2\left[\Sigma(\beta){\tt u}(\tau)+
2{\tt u}(\tau)\right],
$$
\be
0=\ldl v=\Lambda^3\left[\Sigma(\beta){\tt v}(\tau)+
3{\tt v}(\tau)\right],
\l{scalarelemontagne}\ee
where $\beta_{ij}=\ldl \tau_{ij}$ is the $\beta$-funcion and
$\Sigma(\beta)$ is the scaling operator
$$
\Sigma(\beta)=
\beta_{11}{\partial\over\tau_{11}}+
\beta_{12}{\partial\over\tau_{12}}+
\beta_{22}{\partial\over\tau_{22}}.
$$
Note that the solution of the system (\ref{scalarelemontagne})
completely determines the $\beta$-function of the theory.

We now derive the exact $\beta$-function 
projected on the natural moduli directions in terms
of the modular invariants 
$$
J_{\alpha\beta\gamma}=\partial_\alpha a^i\partial_\beta\tau_{ij}
\partial_\gamma a^j,
$$ 
which are completely symmetric in their indices.
Actually, defining the projected $\beta$
function $$\beta_{\alpha\gamma}=\partial_\alpha a^i
\beta_{ij}\partial_\gamma a^j,$$ and using (\ref{oidx1}), we have
\be
\beta_{\alpha\gamma}=-2uJ_{\alpha2\gamma}-3vJ_{\alpha3\gamma}.
\l{yrifi}\ee

The $J_{\alpha\beta\gamma}$'s are related to the 
$I_\beta^{\,\,\gamma}$'s by
$$
I_\beta^{\,\,\gamma}J_{\gamma\beta2}={\pi\over 2i},
\qquad
I_\beta^{\,\,\gamma}J_{\gamma\beta3}=0,
$$
that is
\be
{P(u,v,\Lambda)\over u}J_{223}+12uvJ_{233}=0
={P(u,v,\Lambda)\over 3}J_{333}+12uvJ_{233},
\l{odax}\ee
\be
{P(u,v,\Lambda)\over u}J_{222}+12uvJ_{223}={2i\over\pi}=
{P(u,v,\Lambda)\over 3}J_{233}+12uvJ_{223}.
\l{sysca}\ee
Inserting the solution of this system in (\ref{yrifi}), we obtain 
\be
\beta_{22}={2Au\over 3}[P(u,v,\Lambda)-54v^2],
\l{b1}\ee
\be
\beta_{23}=\beta_{32}={3Av\over u}[P(u,v,\Lambda)-8u^3],
\l{b2}\ee
\be
\beta_{33}=2A[P(u,v,\Lambda)-54v^2],
\l{b3}\ee
where $A= {2i\over \pi}[(12uv)^2-P^2(u,v,\Lambda)/3u]^{-1}$.

\vspace{1cm}

\noindent
{\bf 7.} Let us make some  remarks. 
First of all we note that similar structures 
can be generalized to the case of gauge group $SU(n)$, $n>3$.
Furthermore the condition ${\tt s}(\tau)=0$ 
and the WDVV
equation suggest a relation
with the condition on the lattices obtained in \cite{mmm}
(see Eq.(5.22) there). In this framework one should be able to connect 
the BPS mass formula with the area of degenerate metrics 
on a suitable Riemann 
surface. This surface should be related to the two-dimensional space
which arises in
 compactifying $N=1$ in $D=6$ to obtain $N=2$ in $D=4$.
In \cite{mm} a similar structure 
for the $SU(2)$  case has been obtained.

We also observe that the way we use the Picard-Fuchs 
equations could be useful in investigating some algebraic-geometrical
structure and some aspects concerning mirror 
symmetry (see \cite{varie} for related aspects).
In this context we note that considering the branching 
points of the hyperelliptic Riemann surfaces 
as punctures on the Riemann sphere it should be possible
to describe the Seiberg-Witten moduli space in terms of moduli 
space of Riemann spheres with punctures. Observe that already 
in the $SU(2)$ case
the moduli space is the Riemann sphere with three punctures
which can be essentially seen as 
$\overline {\cal M}_{0,4}$,
 the moduli space of Riemann
spheres with four punctures. In this framework one can use
relevant structures such as the Deligne-Mumford compactification
$\overline {\cal M}_{h,p}$, where ``punctures never collide'',
which allows us to consider natural embeddings (this problem
is of interest also for softly supersymmetry breaking \cite{ADKM}). 
We also observe that
the WDVV equation can be seen as an associativity condition for divisors
on $\overline {\cal M}_{0,p}$ \cite{Kontsevich}.
These structures together with the restriction phenomenon of the
Weil-Petersson metric, whose K\"ahler potential
is the on-shell Liouville action, are at the basis of recursion relations
arising in 2D quantum gravity.

In conclusion, we have found nonperturbative relations for $N=2$ SYM 
with gauge groups $SU(3)$ which generalize the results in \cite{m} where 
the relation between $u$ and the prepotential has been found in the 
$SU(2)$ case. This relation has been recently verified in 
\cite{FucitoTravaglini} up to two-instanton and at all orders in 
\cite{DoreyKhozeMAttis,HoweWest}. The results of our investigation
should be similarly verified for a more complete proof of the 
Seiberg-Witten theory.

\vspace{1cm}

\noindent
{\bf Acknowledgements}. It is a pleasure to thank J. De Boer, B. De Wit,
N. Dorey, A. Faraggi, F. Fucito,  V.V. Khoze,
K. Landsteiner, K. Lechner, W. Lerche, 
P.M. Marchetti, M.P. Mattis, 
P. Pasti, S. Shatashvili, S. Theisen,
M. Tonin and F. Toppan for discussions.

\end{document}